\documentstyle{article}

\begin{document}

\begin{center}
{\huge\bf Foundation of The Two dimensional Quantum Theory of Gravity}
\end{center}

\vspace{1cm}
\begin{center}
{\large\bf
F.GHABOUSSI}\\
\end{center}

\begin{center}
\begin{minipage}{8cm}
Department of Physics, University of Konstanz\\
P.O. Box 5560, D 78434 Konstanz, Germany\\
E-mail: ghabousi@kaluza.physik.uni-konstanz.de
\end{minipage}
\end{center}

\vspace{1cm}

\begin{center}
{\large{\bf Abstract}}
\end{center}
\begin{center}
\begin{minipage}{12cm}
The two dimensional substructure of general relativity and gravity,  
and the two dimensional geometry of quantum effect by black hole  
are disclosed. Then the canonical quantization of the two  
dimensional theory of gravity is performed. It is shown that the  
resulting uncertainty relations can explain black hole quantum  
effects. A quantum gravitational length is also derived which can  
clarify the origin of Planck length.
\end{minipage}
\end{center}

\newpage

{\Large 1. Introduction}

In this report we will point out the main role played by the two  
dimensional substructures of the four dimensional Riemannian space  
time manifold, i. e. the curvature two form and the two dimensional  
subspace of motion or orbits, in the classical- as well as in the  
quantum gravity.

We show in the following that not only the gravitational effects  
are two dimensional effects, but also the "four dimensional" theory  
of gravity has a two dimensional origin. Accordingly we will  
conjecture a two dimensional model for gravity which is defined on  
this subspace of motion. In other words we describe the motion or  
dynamics of a gravitional system just in the motion space or the  
dynamical manifold which is constituted from dynamical degrees of  
freedom of the system. Note that the space of motion is embeded in  
the space-time manifold, but is not identical with this. Thus we  
consider only the actual or dynamical degrees of freedom of the  
object under gravitation field, which is equal to the two dimensions  
of the surface where the motion of such an object takes place. If  
one of such two dynamical degrees of freedom is considered as time  
dependent, then one may speak of one space- and one time coordinate  
or degree of freedom. Nevertheless in accord with the phase space  
philosophy the time is not considered in our model as a new  
independent variable. Therefore in general the space of motion in  
our model is the surface of motion which is described like any  
surface by two variables, this surface is the dual object to the  
gravitational curvature which causes the gravitational motion under  
consideration (see also below).

We will show that, since all gravitational effects including the  
quantum effect by black holes \cite{haw} are two dimensional  
effects, i. e. they depend on maximally two independent variables,  
therefore a two dimensional theory is adequate to describe the  
gravity. Thus such a two dimensional theory has the advantage that  
it is quantizable, despite of the non-renormalizable four  
dimensional general relativitistic theory (GRT) of gravity.

The main point about the geometrical description of gravity with  
respect to the curvature of space around a gravitional source, which  
becomes clarified later in detail, is the following:

In view of the fact that in accord with the general definition of  
curvature by Gauss, the total curvature or the surface integral of  
curvature is given as the excess of the sum of the angles of a  
triangle, on the surface, from $2 \pi$, i. e.: $\phi = \delta \Phi =  
\phi - 2 \pi = \int_{(surface)} R_{ij} dx^i \wedge dx^j$. Therefore  
since the surface and triangle are two dimensional concepts, hence  
the curvature of a manifold is always a two dimensional quantitiy,  
irrespective of the dimension of that manifold. Thus the curvature  
tensor of any manifold has dimension $L^{-2}$. In this sense all  
effects of gravity, i. e. the effects of the curvature of the  
manifold arround the gravitational source, are of two dimensional  
origin, irrespective of dimension of that manifold.

Note that it is the inherent and invariant two dimensional  
property, i. e.  the $L^{-2}$ character of the curvature tensor  
$R_{[ij]}$, which results in the mentioned two dimensional features  
of gravity: Since the dimension of the tensor components of a  
curvature two-form remains always $L^{-2}$ independent of the  
dimensions of the manifold where it is defined \cite{a}. This means  
but that only two independent variables or components are needed two  
define such a curvature, thus also only two independent variables  
are needed to define the dual surface. Thus irrespective of  
dimensions of the underlying manifold where the curvature form and  
its action-invariant are defined, the resulting equations of motion  
for the curvature components are constituted so that these invariant  
character of two dimensionality $\sim L^{-2}$ remains unchanged. In  
other words in view of the fact that the action invariant respects  
the invariant properties of its constituents, i. e. of its  
Lagrangian content and of its integration manifold. Therefore the  
rest components of a curvature tensor on a higher dimensional  
manifold, which is involved in an action invariant on such a  
manifold, are related by the equations of motion in such a manner  
that only two independent components remain. This is the same what  
one confirmed by the two independent components of the metric field  
in the gravitational waves, although one started there from the four  
dimensional Einstein theory.

Thus in the same way that gravitational effects like the planetary  
motion take place on the two dimensional projective surfaces, also  
one needs to describe them only by two degrees of freedom on such  
surfaces, since these are described also by only two independent  
variables. Thus degrees of freedom of a dynamical system are  
equivalent to the independent variables of the system.

In other words to describe dynamical gravitational effects one  
needs to describe the motion orbits of the object under  
gravitational force. However in accord with the $L^{-2}$  
dimensionality of the gravitational curvature or in accord with the  
$L^{-1}$ dimensionality of the gravitational potential, these orbits  
are always two dimensional curves like ellipse or circle which can  
be desribed by only two independent variables ( see Kepler problem  
or perihelion shift, respectively). In this respect these two  
variables and the two degrees of freedom of the moving object under  
gravitational force are synonyms, since the geometrical concept of  
dimension is related with the dynamical concept of degrees of  
freedom in accord with the geometrical interpretation of  
gravitational force by the curvature.

In this sense the two dimensional theory should be formulated on a  
two dimensional {\it curved} submanifold of the three dimensional  
space, where all the gravitational effects including the planetary  
motion, take place. Thus in accord with the geometrical equivalence  
between the gravitational force and the curvature which are both of  
dimension $L^{-2}$ in geometric units, the mentioned two dimensional  
curved submanifold of gravitional effects can be considered as the  
curved surface arround the gravitional source. Accordingly, this  
curved surface $(\sim L^2)$ and the gravitational curvature $(\sim  
L^{-2})$ can be considered as "dual" aspects of the same gravity.
Thus the integral of the two dimensional curvature $(\sim L^{-2})$  
over this two dimensional surface $(\sim L^2)$ which is the  
topological invariant of the surface manifold, is the action  
invariant of our two dimensional model.

The paper is organized as follows: In the second section we  
disclose the fundamental two dimensional substructure of classical  
and quantum gravitational effects as well as of the four dimensional  
theory of gravitation. In sect. 3 we introduce the two dimensional  
classical and quantum theory of gravity. The forth section contains  
the conclusion on the classical limit and the quantum structure of  
space-time. In the appendix we apply the Hirzebruch-Riemann-Roch  
theorem to the mentioned two dimensional theory of gravity in order  
to demonstrate the relevance of its geometric quantization.

{\Large 2. The two dimensional substructure of gravity}

\bigskip
There are various classical- and quantum hints about the main role  
played by the underlying two dimensional spatial substructure (2D)  
of the space-time four manifold in the gravitational effects, in  
Einstein-Hilbert action and in the original approach to quantum  
effects by black holes (BH) \cite{haw}. However, the dominance of  
four dimensional relativistic point of view in classical physics and  
the hierarchy of local point of view in quantum physics, i. e. the  
hirarchy of differential equations, overrun these hints.

Note that all the following hints for the two dimensional  
background of gravity can be considered also as the advantages of a  
two dimensional model of gravity.

The main physical hint about the two dimensional character of  
grivity is the inherent and exclusive two dimensional structure of  
all measurable gravitational effects and the fact that only certain  
two body problems are exactly solvable \cite{KAM}. Thus such a two  
body problem can be reduced to a " single body" problem with two  
degrees of freedom or to the relative motion of a single body with  
the reduced mass.
Note that as the planetary motion in the Newtonian theory, also all  
"general relativistic" gravitational effects, i. e. the perihelion  
shift, the bending of light, the precession of gyroscope and the  
redshift, are two dimensional effects, or a (1+1)-dimensional effect  
which is equivalent two a two dimensional one: The first three are  
describable as the curvature or "angle"-effects on the surface of  
motion of test body in a gravitational field, where the angle which  
manifests any of these effects, i. e. $\Delta \phi \sim  
\displaystyle{\frac{M}{r}}$, is given by the {\it surface} integral  
of the curvature two form: $\phi \propto \int\limits_{(surface)}  
\bar{R}$ \cite{nnnn}. The redshift is a (1+1)-dimensional effect,  
since in this effct only the frequence $( \sim (T)^{ -1} )$ and the  
radial coordinate of Schwarzschild metric $(\sim r)$ are relevant.  
Note that a two dimensional effect is an effect which is of order  
$\displaystyle{\frac{1}{r}} \sim L^{-1}$ or  
$\displaystyle{\frac{1}{r^2}} \sim L^{-2}$ in geometric units. Thus  
it can be described within a two dimensional system which is  
described by the action $ W_G = \phi \propto \int\limits_{(surface)}  
\bar{R} = \oint\limits_{(contour)} \Gamma \ \ $ in accord with  
Stokes theorem, where $\Gamma$ is the gravitational connection form.
Note also that, in view of $M \sim L$ dimensionion in general  
relativistic units, the usual measurable value $\Delta \phi \propto  
\displaystyle{\frac{M}{L}} \sim \displaystyle{\frac{L}{L}}$ is in  
accord with $\phi \propto \int\limits_{(surface)} \bar{R}$, from  
which one obtaints: $\Delta \phi \propto \bar{R} \cdot \Delta A \sim  
L^{ -2} \cdot L^2$, where $A$ is the area of the involved surface.

The second hint is already obvious from the structure of  
Einstein-Hilbert (E-H) action $\kappa^{-1} \int d^4 x \sqrt{- g}  
\tilde{R}$, where in view of $L^{-2}$ dimensionality of curvature  
scalar $\tilde{R}$ and $L^4$ dimensionality of four volume,

($\kappa \sim G$) is forced to be of dimension $L^2 (\sim l^2  
_{Pl})$. Whereby it is just this circumstance which leads to the  
non-renormalizability of quantized (E-H) action \cite{wit2}. Thus  
the four dimensionality of volume is retained here only with the  
help of $L^{2}$ dimensionality of $\kappa \sim G$ which prevents  
however a renormalizable quantization of action. From this hint one  
can conclude (carefully) that a renormalizable quantization is in  
some sense related with a $2D$ volume, since the $L^{- 2}$ dimension  
of disturbing term compensates just the $L^2$ part of four volume.  
Further note that the scalar $\tilde{R}$ is costructed by mixed  
traces between the tangential and curved indices of the curvature  
tensor. Therefore also this action is based on the curvature two  
form which can be defined, as a two form, entirely on a two  
dimensional manifold.

The third hint for the two dimensionality of gravity comes from the  
theoretical analysis of the gravitational waves. It is known that  
these waves, despite of their four dimensional background, are  
transversal and possess only {\it two} degrees of freedom which  
gives the dynamical degrees of freedom of the metric field  
\cite{nnn}. In view of the fact that the linearization of theory,  
which enables one to derive these waves, does not affect the number  
of degrees of freedom of the theory, also the assumed full  
"non-linear" gravitational waves or the dynamical gravitational  
metric field possess only two degrees of freedom. This is not only  
in accord with the related concept of square of length or area which  
is a two dimensional invariant, but it is also in accord with the  
number of degrees of freedom of the curvature form which can be  
considered as the second derivitive of the Riemannian metric. Thus  
in view of the fact that the number of independent degrees of  
freedom of the metric field determines the number of independent  
components of this field and this depends on the dimension of the  
underlying manifold, hence the dimension of the underlying manifold  
of such a metric should be also two. This is the two dimensional  
submanifold of the four dimensional spac-time manifold which is  
involved in the gravitional dynamics that can be described either  
with respect to the metric tensor or with respect to the related  
curvature form.

The forth hint comes from the original approach where the BH  
entropy is given by

$S = (constant) \cdot \displaystyle{\frac{1}{4}} A_{(3D)}$ with $A$  
the area of BH event horizon(s). Now the area of $S^2$ which is the  
surface of general topological $3D$ object, i. e. the sphere, is  
$4\pi r^2$. One quarter of which is just the area of its cross  
section, a disc, which is the general topological $2D$ object.  
Therefore, the entropy $S$, as a quantum theoretical quantity,  
should be considered as to be proportional to the area of the $2D$  
submanifold and not to the area of a classical $3D$ object, i. e. it  
should be $S := (constant) \cdot A_{(2D \subset 3D)}$ (see below).
Thus the proportionality between entropy and area even in the four  
dimensional approach in view of the fact that the area is a two  
dimensional invariant, shows the fundamental role of $2D$ structure  
within the quantum structures on four manifolds.

The fifth hint for the two dimensional structure of the GRT of  
gravity and of black holes comes directly from the analysis of  
singularity in GRT which was lead to the concept of black holes:  
Although the BH is considered as the singularity of the four  
dimensional Schwarzschild solution of Einstein equations,  
nevertheless the singularity analysis shows that such a four  
dimensional solution is not suitable to determine the true  
singularities.

First note that the "four dimensional" Schwarzschild solution in  
the spherical coordinates is in fact a three dimensional solution,  
since the relation $x^2 + y^2 + z^2 = r^2$ in spherical coordinates  
reduces the number of independent spatial variables from three to  
two. Thus $r$ is here not a new variable, but a function of $(x, y,  
z)$ and the $x^2 + y^2 + z^2 = r^2$ is the equation of a spherical  
{\it surface} which can be described also by only two independent  
variables. Moreover in the so called extended solutions, i. e. in  
Lemaitre, Eddington-Finkelstein and Kruskal metrics, one reduced  
these three variables by further linear combination to only two  
independent variables. Thus in the phase space description of  
Einstein equations, any spherical, i. e. any three dimensional  
solution is in fact a two dimensional solution, since the time is  
not a true independent variable in the phase space, but a parameter.  
In other words even the so called "four dimensional" solutions of  
Einstein equations are in fact two dimensional ones. This is not  
surprising, since in accord with the above discussion the variable  
part of the E-H action comes from a two dimensional curvature tensor  
$( \sim L^{-2} )$.

Thus it is known that just the extended two dimensional Kruskal  
metric \cite{kr} is the metric which delivers the true singularity  
in $r = 0$, whereas the "four dimensional" Schwarzschild metric  
contain false singularity in $r = 2 M$ which is related with  
coordinate effects. Note that this fact is a hint about the two  
dimensional background of BH solutions of Einstein equations. Again  
in view of the fundamental relation between the dimensions of metric  
and of the underlying Riemannian manifold, the underlying manifold  
with such a two dimensional metric  is also of dimension two.
Therefore the responcible theory of gravity for such a "two  
dimensional" singularity, which is a dynamical theory of the  
underlying manifold, should be two dimensional theory. In other  
words black hole as singularity is a two dimensional concept and in  
this sense it is also a concept of two dimensional theories. Thus in  
four dimensional theories with four dimensional metrics one needs a  
reduction to the two dimensional metric to come close to the true  
singularities. It is also importent to mention that in view of the  
fact that singularities belong to the invariant properties of a  
theory, therefore if one will replace a theory, then the new theory  
should contain the same singularities as the old one. Hence even in  
this sense the two dimensional theory of gravity is the best  
candidate to replace the four dimensional theory.

\bigskip
As the last classical hint in favour of a two dimensional approach  
let us mention the constraint disadvantage of four dimensional  
theories, e. g. the GRT. Such constraints arise from the  
superficiality of the {\it time component} of dynamical variable of  
the theory, which however can not be a dynamical {\it component} in  
view of the canonical or symplectic structure of phase space of  
theory. Note that the time is not a true variable in the phase space  
of a physical system. Thus in the so called extended phase space of  
a one dimensional motion, where momentum and position are  
considered as functions of time parameter \cite{wood}, time plays  
the role of "canonical conjugate" {\it parameter} to the Hamiltonian  
which is itself a function of momentum and position variables.  
Hence we have to do even in the extended phase space with two  
interdependent pair of independent canonical conjugate variables, i.  
e. momemtum, position or time and Hamiltonian, from which only one  
pair are independent physical variables.
Related with this disadvantage is the problem of a covariant  
definition of energy-momentum complex in the four dimensional  
theories, which becomes more critical with respect to the  
quantization. Thus in the GRT one can not define a covariantly  
conserved quantity for the energy of gravitational field.

Note however that the discussed two dimensional substructure of the  
four dimensional space-time does not affect the general relativity  
between space and time "coordinates". Thus all measurable  
consequences of such a relativity can be described with respect to  
one space and the time coordinates, which is equivalent to a two  
dimensional effect in view of the theory of relativity where $T \sim  
L$.

These hints disclose the two dimensional substructure of the four  
dimensional theory of gravity. This is done in the sense that we  
showed that although Einstein theory and its solutions for gravity  
are formulated in a four dimensional fashion, however since they are  
based on the {\it two dimensional} concept of curvature, therefore  
they possess only two truely independent dimensions or two degrees  
of freedom. A fact which is in accord with the two dimensional  
nature of gravitational effects. The rest two dimensions of this  
four dimensional theory are frozen in the dimensional coupling {\it  
constant} of the theory $\kappa$, so that they play no dynamical  
role.

\bigskip
Now we discuss a quantum hint which comes from the quantum effect  
by black holes. A geometrical analysis of the original approach to  
this effect \cite{haw} shows that also the underlying model is a two  
dimensional model with respect to the quantum effect of particle  
creation by BH. Since the responcible quantum effect in BH is a  
"Aharonov-Bohm" like {\it gravitational} effect, whereby the  
gravitational curvature field plays the same role as the magnetic  
field in Aharonov-Bohm effect. Note that even the Aharonov-Bohm  
effect is a two dimensional effect in the sense that the electron  
moves on a two dimensional surface which contains a cross section of  
the solenoid. Thus the phase change of electron is given by a  
contour integral of the electromagnetic potential which surround  
this surface. Moreover, it is a typical global quantum phase effect  
\cite{typ} which rises in the original approach just by a global  
comparison between quantum operations in two spatially separated  
regions which are asymptotically flat. Thus this asymptotic flatness  
of outside region of BH, with respect to the curvature of  
space-time, is comparable with the vanishing of magnetic field  
outside of solenoid in Bohm-Aharonov effect. Hence from quantum  
mechanical point of view the Hermitian scalar quantum field $\Phi$  
receives a phase surplus, i. e. a change of frequence, which is  
caused in the original approach by the comparison of anihilation  
operations in the two asymptotically flat regions in view of the  
imposibility of an invariant definition of positive and negative  
frequences under the influence of the curved BH metric [1]. Then  
this {\it global} comparison can be considered, as we will show, as  
a looping around the strong gravitational field of BH, where the  
looping takes place in view of comparision of "operations" in the  
two asymptotically flat regions around the BH in the following way:

First recall that in the original approach where $\Phi = \Sigma_i  
{\{ f_i a_i + \bar{f}_i a^{\dagger}_i }\}$ the operators $a_{1 i}$  
and the quantum state $|0_1>$ are defined in the asymptotically flat  
region (1), whereas the operators $a_{3 i}$ and $|0_3>$ are defined  
in the asymptotically flat region (3) with the BH region (2)  
between them. Therefore, the operation $a_{3 i} |0_1>$ can be  
considered as if $|0_1>$ is moved to the region (3), where $a_{3 i}$  
is defined, to experience this operation and the subsequent  
comparison of $a_{3 i} |0_1>$ with $a_{1 i} |0_1>$ which is defined
in region (1) can be considered as if $|0_1>$ is moved back from  
region (3) to its original region (1). The quantum effect rises then  
by this global comparison between operations of anihilation  
operators on vacuum state in two distant regions (1) and (3). In  
this sense one can consider the above comparision of operations on  
the vacuum state $|0_1>$ as a looping of state $|0_1>$ around the  
strong gravitional region (2) from region (1) to region (3) and back  
to the region (1). Obviously one can consider instead of that also  
the loop of operator $a_{3 i}$ from region (3) to region (1) for  
operation $a_{3 i} |0_1>$ and back to the region (3) where the  
quantum effect rises by comparison of $a_{3 i} |0_1> \neq 0$ with  
$a_{3 i} |0_3> = 0$.
Hereby the quantized flux of gravitional curvature field in region  
(2) plays the similar role as the role played by the flux of  
magnetic field in Bohm-Aharonov effect. Therfore even the four  
dimensional original model to describe quantum effects by BH is  
properly a two dimensional model on the loop surface of vacuum state  
of $\Phi$ which is similar to the two dimensional Bohm-Aharonov  
model on the motion surface of electrons around the cross section of  
solenoid. In this sense the discussed quantum effect by BH is again  
a "phase" or "angle" effect which is given by the surface integral  
of the curvature two form.

Thus one can conclude that not only the classical gravitational  
effects but also the quantum gravitational effect by black hole are  
two dimensional effects which are caused by the gravitational  
curvature tensor in accord with the action $W_G = \phi \propto  
\int\limits_{(surface)} R_{mn} dx^m \wedge dx^n =  
\oint\limits_{(contour)} \Gamma_m dx^m \ \ ; \ \  m,n = 1, 2$.  
Therefore we conjecture in next section that the gravitational  
action functional should be given, up to a {\it dimensionless}  
coupling constant $\tilde{G}$, by this action. Such an action  
describes the gravitational flux through the surface which is  
bounded by the contour orbit of a moving object under the  
gravitational influence. This is in the same manner as the magnetic  
flux perceived by the electron moving around the magnetic field.

Note that  there is an intrinsic merit of the two dimensional model  
with respect to the four dimensional theory, since the length or  
distance $x_m$ is a canonical variable in the two dimensional model  
in accord with the fact that the canonical conjugate variables of  
this action are ${\{ \Gamma_m  \ \ , \ \ x^m }\}$. Therefore the two  
dimensional quantized model possess a quantum of length which is  
given in accord with the quantization of the action in canonical  
manner: $W_G = \tilde{G} \int\limits_{(surface)} R_{mn} dx^m \wedge  
dx^n = \oint\limits_{(contour)} \Gamma_m dx^m = N h ; \ N \in  
{\mathbf Z}$ (see next section). Such a quantum of length is given  
in our model by: $l_G ^2 = \displaystyle{\frac{\hbar}{\tilde{G} R}}$  
similar to the magnetic length in magnetic quantization  
\cite{aoki}. Here $R := \epsilon_{mn} R_{mn}$ is the constant  
gravitational curvature arround the source, since in two dimensional  
case the curvature is constant. Thus we can define and speak of a  
quantum of length like the Planck length in the two dimensional  
model. Whereas since the length is no canonical conjugate variables  
in the four dimensional GRT and GRT is not quantizable, one can not  
define a quantum of length like Planck length in GRT or even in any   
hypothetical quantization of GRT.

Summarizing the above discussions the advantages of two dimensional  
model with respect to the four dimensional GRT of gravity are  
manifolds: Among them: The two dimensional model avoids the  
coordinate effects which appear in the four dimensional theory with  
respect to the singularities. Despite of constraints in the four  
dimensional theory which results in ambiguities in quantization of  
this theory, the two dimensional model possess no constraint. Thus  
the two dimensional model is canonically quantizable, whereas the  
four dimensional theory is not quantizable in view of its  
dimensional constant $\kappa$, its metrical structure and  
constraints. The two dimensional model is more appropriate for  
gravity in view of the two dimensional structure of gravitational  
effects including quantum effects by black holes. Thus the two  
dimensional model is formulated directly in the motion space where  
the gravitational effects take place.

{\Large 3. The two dimensional classical and quantum theory of gravity}

The main reason for the emergence of quantum effects in two  
dimensions is that quantum effects are invariant (global) effects  
which are based principielly on the existence of a quantum structure  
on the phase space of physical system which itself is based on the  
existence of a flat complex line bundle $(\sim U(1)_{flat})$ over  
the phase space. Thus the reason that such a two dimensional  
quantization can be responsible for quantum effects, which are  
originally formulated for a four dimensional theory, is that the  
quantization of a system takes place on the phase space of system  
and not on the space-time where the system is usually defined. In  
other words the number of dimensions of space-time has no influence  
on the possibility that a system becomes a quantum system, but only  
on the structure of its quantization \cite{NN}. Whereas the  
possibility that a system becomes a quantum system depends on the  
value of its action $W$, i. e. if $W \sim \hbar$ \cite{feyn}.

Moreover the mentioned flat principal $U(1)$ bundle is closely  
related with the symplectic structure of phase space of syste:  
$\omega := d \pi_m \wedge dq^m$ \cite{wood}. Therefore we give first  
the symplectic structure of the two dimensional model:

The curvature two form $\bar{R} = R_{mn} dx^m \wedge dx^n$ admits a  
symplectic structure on the two dimensional surface $M$, i. e.  
$\bar{R} \sim \omega = d \pi_m \wedge dq^m$, in view of its  
closedness $d \bar{R} = 0$ and its non-degeneracy $R_{mn}  
^{(surface)} := constant  \neq 0$. Hence $q^m \sim x^m$ and $\pi_m  
\sim \Gamma_m = R_{mn} \cdot x^n$. In this sense the two dimensional  
classical action of our two dimensional model for gravity: $W_G =  
\oint_{(contour)} \Gamma_m dx^m = \int\limits_{(surface)} R_{mn}  
dx^m \wedge dx^n$ is equivalent to the symplectic action function  
$\int\limits_{phase \, space} d \pi_m \wedge dq^m$ on the phase  
space of this two dimensional gravitational system with its  
canonical conjugate variables ${\{ R_{mn}x^n \, ,\, x^m }\}$. This  
is again equivalent by $\int\limits_{phase \, space} d \pi_m \wedge  
dq^m  = \oint\limits_{phase \, space} \pi_m dq^m$ to the integral  
$\oint\limits_{phase \, space} \pi_m dq^m$ on the same phase space  
with the equivalent canonical conjugate variables ${\{ \Gamma_m \, ,  
\, x^m }\}$ (see also below).

>From now on we consider the curved surface of integration $( M ;  
\bar{R} )$ as a symplectic manifold with respect to the curvature  
structure $\bar{R}$. Hence in this sense the symplectic manifold $(  
M ; \bar{R} )$ can be considered as the phase space of {\it two  
dimensional} gravitational system.

This idendification of symplectic structure on $(M ; \bar{R})$ is  
in principle enough to consider a canonical quantization on $(M ;  
\bar{R})$ by postulating $W_G = N h$ as usual.
Thus for physically reasonable situations, e. g. for strong  
gravitational curvature where $W_G \sim \hbar$, it can be expected  
that we meet the quantization of such a symplectic structure.

One obtains the two dimensional equations of motion for $\Gamma_m$  
from the variation of this action with respect to the variation of  
$x^n$. Here, i. e. on the surface, $\Gamma_m$ depends also on $x^n$  
so that one should consider the variation of $W_G$ only with respect  
to the variation of $x^n$ whereby $dx^m =  
\displaystyle{\frac{\partial x^m}{\partial x^n}} dx^n$. Then the  
Euler-Lagrange equation $\displaystyle{\frac {\partial L}{\partial  
x^m}} = \partial_n {\frac {\partial L}{\partial \partial_n x^m}}$  
results in the equations of motion $\partial_n \partial^n \Gamma_m =  
(d^{\dagger} d + d d^{\dagger}) \Gamma = 0$
which is the usual Laplace equation of motion for $\Gamma_m$ in two  
dimensions \cite{nn}.
Note also that in the two dimensional case under consideration,  
where $\bar{R}$ is a constant two form, the equations $d \bar{R} =  
0$ and $d^{\dagger} \bar{R} = 0$ which are equivalent with the above  
equations of motion \cite{nn}, are identities. In this sense the  
equations of motion in the two dimensional case are purely  
geometrical statements.

\bigskip
Note that, in view of $W_G = \oint_{(contour)} \Gamma_m \dot{x}  
dt$, the canonical conjugate variables of the phase space of this  
two dimensional gravitational action are given in accord with the  
Legendre formula $\displaystyle{\frac {\partial L}{\partial \dot{x}  
^m}}$, by: ${\{ \Gamma_m \, , \, x^m }\}$ variables. Hence we can  
potulate the canonical quantization of the related system directly  
by $W_G = N h$. Nevertheless we show in the appendix that such a  
canonical quantization is even in accord with the general structures  
of geometric quantization.

\bigskip
To perform the canonical quantization of two dimensional  
gravitational system recall that the canonical quantization of  
action $W := \int \limits_{(phase \, space)} \omega = \int  
\limits_{(phase \, space)} d \pi_a \wedge dq^a$ can be described by  
the equivalent postulates: $W = N h \, , \, N \in {\mathbf Z}$ or  
$[\hat{\pi}_a \ ,\, \hat{q}^b ] = -i \hbar \delta_a ^b$. Therefore,  
in view of the fact that the symplectic manifold $( M ; \bar{R} )$  
plays the role of phase space of our two dimensional gravitational  
system, one can expect that it can be canonically quantized, if the  
value of gravitational action is of order $\hbar$ \cite{feyn}. In  
this case the two dimensional gravitational system can be considered  
as a quantum system on the compact orientable two dimensional  
Riemannian manifold $M$ without boundary and its quantization can be  
postulated by

\begin{equation}
W_G ^Q = \int \limits_{surface} R_{mn} dx^m \wedge dx^m = \oint  
\limits_{contour} \Gamma_m dx^m = N h
\end{equation}

Note also that we have to do here with a non-simply connected  
region which contains of inside curved region of BH (surface) $M$  
and the outside asymptotically flat (contour) region, where  
curvature $\bar{R}$ and connection $\Gamma$ have different values.  
Thus $R^{mn} _{(surface)} = constant$ and $R^{mn} _{(contour)} = 0$,  

whereas $\Gamma^m _{(surface)} = R^{mn}_{(surface)} \cdot x^n$ and  
$\Gamma^m _{(contour)} = \Gamma^m _{(flat)}$ with $ d \Gamma^m  
_{(flat)} \equiv 0$. Note that although $\Gamma^m _{(contour)} =  
\Gamma^m _{(flat)}$ can be locally gauged away, nevertheless $\oint  
\limits_{contour} \Gamma_m dx^m \neq 0$ is an invariant of the  
system which is equal to the value of action functional of system  
$W_G = \int\limits_{surface} R_{mn} dx^m \wedge dx^n$  
\cite{erkBAme}.

Therefore the canonical quantization of this two dimensional  
gravitational action can be performed either on the phase space of  
contour region with canonical conjugate variables ${\{ \Gamma_m \, ,  
\, x_m}\}$ in accord with $[ \hat{\Gamma}_m \ , \, \hat{x}_m ] = -i  
\hbar$ or it can be performed on the equivalent phase space of the  
surface region with canonical conjugate variables ${\{ R_{mn}.x_n \,  
,\, x_m}\}$ in accord with $R [ \hat{x}_m , \hat{x}_n ] = -i \hbar  
\epsilon_{mn}$. The equivalent uncertainty relations are then given  
by, respectively:

\begin{equation}
\Delta \Gamma_m \cdot \Delta x_m \geq \hbar \ \   ,  \ \ R \cdot  
\Delta A = R \cdot | \epsilon_{mn} | \Delta x_m \cdot \Delta x_n   
\geq \hbar\ \ ,
\end{equation}

where $\Delta A$ is the area uncertainty.

Thus the quantum commutator postulate $[ \hat{\Gamma}_m \ , \,  
\hat{x}_m ] = -i \hbar$ and the equivalent quantization

$\oint \limits_{contour} \Gamma_m dx^m = N h$ are comparable,  
respectively, with the canonical quantization postulates:

$[ \hat{\pi}_m \ , \, \hat{q}_m ] = -i \hbar$ and $\oint  
\limits_{phase \, space} \pi_m dq^m = N h$.

Moreover the operators $\hat{\Gamma}_m$ and $\hat{x}_m$ are given  
in the $\Psi_G (x_m , t)$- or $\Psi_G (\Gamma_m , t)$ representation  
of the wave function of quantized gravitational system,  
respectively, by ${\{ \hat{\Gamma}_m := -i \hbar  
\displaystyle{\frac{\partial}{\partial x^m}} \ and \ \hat{x}_m :=  
x_m }\}$ or by ${\{ \hat{\Gamma}_m := \Gamma_m \  and \  \hat{x}_m  
:= i \hbar \displaystyle{\frac{\partial}{\partial \Gamma ^m}} }\}$  
\cite{quant}. Thus in both representations the commutator potulate:

\begin{equation}
[ \hat{\Gamma}_m \, , \, \hat{x}_n ] = -i \hbar \delta_{mn} \ \ ,
\end{equation}

is fulfilled.

However the relative weakness $10^{-40}$ of gravitational  
interaction with respect to electromagnetic interaction sets limits  
on the values of curvature and area of a BH in a quantum state. Thus  
in wiew of such a relation a {\it dimensionless} gravitaional  
coupling constant $\tilde{G}$ can be introduced by $W_G :=   
\tilde{G} \int \limits_{surface} R_{mn} dx^m \wedge dx^m = \tilde{G}  
\oint \limits_{contour} \Gamma_m dx^m = N h$ in analogy with the  
electric charge $e$. Obviously such a coupling constant should match  
with the Newtonian constant in the classical limit of our quantized  
gravitational system. Nevertheless, if not necessary, we set in the  
rest $\tilde{G} = 1$.

\bigskip
Analysing the geometrical structure of this two dimensional quantum  
gravity in accord with

$\Delta \Gamma_m \cdot \Delta x_m \geq \hbar$ it becomes obvious  
that in view of $\Delta x_m > 0$ the two dimensional quantum  
gravitational system or the quantized curved surface $(M ,  
\bar{R})_Q$ posses no one dimensional boundary in the classical  
sense, i. e. $\partial (M , \bar{R})_Q = \emptyset$. Since the  
contour region of $M$ has in the quantized case a width of $\Delta x  
> 0$ which can not be undercut under quantum conditions. Thus the  
quantum contour area with the undetermined width $\Delta x$ can be  
considered as belonging not to the BH but to the outside region of  
BH.

Furthermore, in view of uncertainty relation $R \cdot \Delta A \geq  
\hbar > 0$ the area uncertainty $\Delta A$ is always positive and  
the BH area is increasing. Thus the entropy uncertainty of BH system  
which is related with gravitional quantum effects is given by  
$\Delta S_G \sim \Delta W_G = R \cdot \Delta A \geq \hbar > 0$.  
Hence the entropy of BH system increases permanently as like as its  
area in accord with quantum gravitional processes.

Moreover, in analogy with the two dimensional quantization of  
electromagnetic systems (see the appendix), one should expect that  
in two dimensional quantum gravitational case there exists a quantum  
of length
$(\Delta x_m)_{(minimum)} := l_G$, which is defined in analogy with  
the magnetic length $(\sim l_B ^2 := \displaystyle{\frac{\hbar}{e  
B}})$ by:

\begin{equation}
l_G ^2 := \displaystyle{\frac{\hbar}{\tilde{G} R}}
\end{equation}

It is the most minimal length which is quantum mechanically  
obtainable under the action of constant curvature $\bar{R}$. In  
other words one can derive a quantum gravitational length $l _G$  
from the obtained uncertainty relation by the "uncertainty  
equation": $R \cdot l^2 _G = \hbar$ or $\tilde{G} R \cdot l^2 _G =  
\hbar$, since $\Delta A \geq l^2 _G$. In this sense if a strong  
gravitational source is in a quantum state, then all length and  
areas arround it can be quantized in the units $l_B $ and $l_B ^2$.

Therefore the quantization of action $W_G :=  \tilde{G} \int  
\limits_{M} R_{mn} dx^m \wedge dx^m = N h$ in view of constance of  
$R_{mn}$ is due to the quantization of area $A_{BH} = N \cdot 2 \pi  
l_G ^2$. Thus the quantum entropy should be related also with the  
quantization of area by $S_G \sim W_G = \tilde{G} R \cdot N \cdot 2  
\pi l_G ^2$. Thus it is a major merit of the two dimensional model  
that by this model one can quantize lengths and areas which is at  
the heart of a quantization of geometrical gravity.

It is also possible to relates this gravitational length with the  
Planck length $l^2 _{Pl} := \displaystyle{\frac{\hbar G}{c^3}}$ by  
$\tilde{G} = \displaystyle{\frac{c^3}{G R}}$. In this sense the  
canonically defined quantum of length should replace the Planck  
length. Note however that only, since in the two dimensional model  
$\Gamma$ and $x_m \sim (length)$ are the canonical conjugate  
variables, therefore the length is quantized by the canonical  
quantization of the model. Whereas in the four dimensional GRT the  
length is no canonical variable and therefore the length can not be  
quantized in GRT. Therefore to speak of a quantum of length or  
Planck length even in a hypotetically quantized GRT is an assuption  
without any quantum theoretical foundation. Thus the canonical  
quantizability of two dimensional model, which results in the  
quantization of length and areas, confirms the consistency of this  
model.

\bigskip
As it is mentioned above in view of the uncertainty relations there  
are amount of momentum $\Delta \Gamma \geq  
\displaystyle{\frac{\hbar}{\Delta x^m}}$ or $\Delta \Gamma_m =  
\displaystyle{\frac{\hbar}{\tilde{G} l_G}}$ which can be considered  
as radiated, i. e. as belonging to the region outside of BH. This  
particle creation is related with the indeterminacy of contour area  
as it is discussed above. Of course one can relate this particle  
creation also with the energy time uncertainty relation which is  
equivalent to momentum position uncertainty relation or to the area  
uncertainty relation as discussed above.

\bigskip
To compare our time independent approch with the original one note  
that as in any time dependent system we can also use the extended  
phase space \cite{wood} where the phase space variables, e. g. ${\{  
\Gamma_m \, , \, x_m }\}$ depend on time parameter.
In accord with the extended canonical action: $ \tilde{W} = \oint  
\pi_m dq^m - \int H dt$ the extended two dimensional gravitational  
action of BH is rewritten by: $\tilde{W}_G = \oint \Gamma_m dx^m -  
\int H_G dt$ where $H_G$ is the Hamiltonian of the two dimensional  
BH system. Hence in accord with the canonical quantization  
$\tilde{W} = N h$ where the uncertainty relations: $\Delta \pi_m  
\cdot \Delta q_m  \sim \Delta E \cdot \Delta t \geq \hbar$ rise,  
then also in the  canonical quantized gravitional system  
$\tilde{W}_G = N h$ the following uncertainty relations should rise:  
$\Delta \Gamma_m \cdot \Delta x_m  \sim \Delta E_G \cdot \Delta t  
\geq \hbar$. In this sense the "free" amount of gravitational energy  
$\Delta E_G \geq \displaystyle{\frac {\hbar}{\Delta t}}$ can be  
transformed within the time $\Delta t$ into the pair production  
energy. Recall however that the equivalent connection $(\sim$  
momentum)- position uncertainty, which also can be responsible for  
the pair production by BH, was already derived by the time  
independent canonical quantization of gravitational system.

Furthermore the following temperature uncertainty $\Delta T$ can be  
calculated from the relation between energy and entropy by $\Delta  
E_G = \hbar \cdot \Delta S \cdot \Delta T$ where $\Delta S =  
(\hbar)^{-1} \cdot \Delta W_G$. In this manner the increase of  
entropy and area of BH are related together and with the particle  
production in accord with $\Delta \Gamma_m \cdot \Delta x_m \sim  
\Delta E_G \cdot \Delta t \sim \Delta S \cdot \Delta T > 0$.  
Therefore, the thermodynamics of BH can be obtained also from the  
two dimensional approach. For example to get a zero temprature in  
the quantum mechanical sense, i. e. to arrive $\Delta T = 0$ one  
needs infinite amount of operations, i. e. infinite amount of  
$\Delta S$. Recall that, as it is mentioned above, $\Delta S \sim  
\Delta A > 0$ are always positive and $\Delta S \sim R_{(constant)}  
\cdot \Delta A$. Moreover since $\Delta A_{(minimum)} = l^2 _G \sim  
l^2 _{Pl}$ and the area is quantized in units of $l^2 _G \sim l^2  
_{Pl}$, hence in view of $\Delta S_G \sim R \cdot \Delta A$ and $S_G  
\sim N \cdot \Delta S_G$ the entropy can be considered as given by  
the quantum number: $S_G \sim N$.

Thus all quantum aspects of BH gravitations can be deduced also from this
type of two dimensional quantum gravities which can be formulated  
also as a time dependent theory.

Furthermore for time dependent variables ${\{ \Gamma_m (t) \, , \,  
x_m (t) }\}$ of an extended phase space, one obtains also a  
conventional equation of motion for a test particle moving in a  
gravitational field:

Considering the action:

\begin{equation}
S_G = \frac{1}{2} ( \oint P_m dx^m + \tilde{G} \oint \Gamma_m dx^m  
) \ \ ,
\end{equation}

for such a test particle with momentum and position coordinates  
$P_m$ and $x_m$, the equations of motion should be given by:

\begin{equation}
\dot{P} _m = - \tilde{G} (\dot{\Gamma} _m + \dot{x} _n \cdot  
R_{mn}) \ \ ,
\end{equation}

which is of Lorentz force type.

Note that the canonical quantization of action (6), which posses  
the canonical conjugate variables: ${\{ \displaystyle{\frac{1}{2}} (  
P_m + \tilde{G} \Gamma_m ) \ \ , \ \ x^m }\}$, results beyond the  
test particle commutatator $[ \hat{P}_m \ , \ \hat{x} _m ] = - i  
\hbar$, also in the above obtained commutator (4) with a constant  
factor $\tilde{G}$. Thus one can obtain the above quantization of  
(1) or (2) also from the quantization of (6).

\bigskip
Note also that there is a relation to a possible $(2+1)$  
dimensional Chern-Simons theory of gravitation which can be given by  
the action $\int\limits_{(2+1)} \Gamma \wedge \bar{R}$. Since in  
our model, in accord with $\Gamma_m = R_{mn} \cdot x^n$, the  
commutators
$R \cdot \epsilon_{mn} [ \hat{x}_m , \hat{x}_n ] = [ \hat{\Gamma}_m  
\, , \, \hat{x}_m ] = -i \hbar$ are equivalent to the connutator $[  
\hat{\Gamma}_m \, , \, \hat{\Gamma}_n ] = -i \hbar R \cdot  
\epsilon_{mn}$ which is the commutator postulate for the  
quantization of the gauged Chern-Simons functional  
$\int\limits_{(2+1)} \Gamma_m d \Gamma_n \epsilon_{mn}$
in the $\Gamma_0 = 0$ gauge. Thus our gauge free two dimensional  
quantum gravity model is related with a {\it gauged} $(2+1)$  
dimensional quantized Chern-Simons gravity. Note however that in our  
model the curvature is in view of the two dimensionality a constant  
one, whereas in the usual Chern-Simons model the curvature is  
constrained to be zero.

\bigskip
{\Large Conclussion: The classical limit and the quantum structure  
of "space-time"}

\bigskip
The conclusion is that in view of our discussion of classical  
gravitational effects and of the two dimensional structure of  
original approach to quantum effect by BH, all these effects can be  
described without lose of generality as two dimensional effects.  
Thus, just in view of general relativistic relation between time and  
space "variables" in the extended configuration space any (1+1)  
dimensional effect is equivalent to a two dimensional effect.
Hence in this sense the classical limit of our two dimensional  
model is straightforward, since all classical gravitational effects  
including the planetary motion can be considered as two dimensional  
effects.

In the same manner the flat space-time appears as the large scale  
limit where the curvature vanishes. Since by vanishing of curvature  
and connection, the motion of test particle which is given by (7)  
becomes rectilinear. Thus even the quantum effect by BH can be well  
described with the above canonically quantized two dimensional  
model.

\bigskip
Furthermore it seems that the event horizon of BH is comparable  
with the contour region in our approach where the integral $\oint  
\limits_{contour} \Gamma$ takes place and which has in quantized  
case $\oint \limits_{contour} \Gamma_m dx^m = N h$ a width of  
$\Delta x_m \geq l_G > 0$. The effect of strong curvature $R$ is to  
bind particles to move in this contour region as like as in a  
potential well which is caused by the strong field of curvature. In  
quantum gravitational case one can expect that the absorbed  
particles by BH gravitation are bound in the contour region of BH  
flowing permanently in this region so that they can not scape from  
this region as long as the gravitational field is stronger than  
possible repulsive fields acting on these particles.

As a last remark let us mention that this model has consequences  
for the possible quantum structure of "space-time" which will be  
discussed in more details elsewhere. Nevertheless the first lesson  
from this model for quantum spaces is that such quantum structures  
are possible in two spatial dimensions. In other words the position  
coordinates of a particle moving in a very strong gravitational  
curvature have to be considered in a quantum theory, in accord with  
the above discussion, as quantum operators which fulfil the  
commutator postulate: $R \cdot \epsilon_{mn} [ \hat{x}_m , \hat{x}_n  
] = -i \hbar$. It is in view of the cyclotron motion of particles  
in a curvature field, which is neccesary for the periodic  
quantization, or in view of the two dimensionality of the curvature  
field ($\sim L^{-2}$), that as usual only the surface coordinates of  
the particle becomes quantized. Nevertheless in view of the fact  
that the time is not atrue variable in the phase space which is the  
space of quantization, the time "coordinate" of a particle can not  
be quantized directly. On the other hand, if the position and  
momentum coordinates in the phase space of a system which should be  
quantized become functions of a time parameter, then it is possible  
to define a Hamiltonian as a function of the phase space variables  
and to quantize the energy integral of the system or equivalently to  
obtain uncertainty relations for the product of time and energy.  
This is an indirect canonical quantization of time which requires  
however, in principle, to introduce a quantum time "coordinate"  
operator of the moving particle which should be in general  
proportinal to an energy derivation in the energy representation of  
the wave function of the system of a single particle in the  
gravitational curvature: $\Psi \propto exp (i \int \int \bar{R} )$.

Recall that although there are various models of quantum  
space-times, but none of them gives a complet canonical quantization  
of space-time with respect to a possible phase space- or action  
functional quantization \cite{x}. Thus any correct quantum  
commutator postulate of space-time variable operators or any  
uncertainty relation for them should be equivalent to the action  
functional quantization. Moreover, in view of the dimensional  
discussion at the begining of this paper and in acoord with the  
mentioned two dimensional structure within the original four  
dimensional approch, it seems that any quantum theory of space-time  
or of quantum space-time effects should be equivalent to some two  
dimensional quantum structure.

The main lesson is that in accord with($\hbar \sim L^0)$ in  
geometrical units: In view of the restriction of all known physical  
quantities to those up to two dimensional ones, e. g.: ${\{  
invariants \sim L^0: (action, charge)}\}$,

${\{ connections \sim L^{-1}: ( momentums, \ enrergies, \  
potentials) }\}$ and ${\{ curvatures \sim L^{-2}: (forces, \ field \  
strengths)}\}$. A resonable quantum relation such as the quantum  
action postulate, the quantum commutator postulate or the  
uncertainty relation should be dimensionless within this two  
dimensional structure.

\newpage
{\LARGE Appendix}

\bigskip
First note that any $2 form$ which is defined on a two dimensional  
manifold like $M$ is by definition a closed $2 form$, since there is  
no $3 form$ on $M$. In this sense all such $2 forms$ may represent  
spatially constant field strengths. Note also that $R_{mn}$ can be  
considered as a constant almost complex strucure on the two  
dimensional manifold under consideration. Nevertheless this  
constance of almost complex structure results then in vanishing of  
Nijenhuis tensor \cite{nak} on the considered two dimensional  
(surface) mamifold, so that $(M ; \bar{R})$ admit a complex  
structure which simulates a $U(1)$ bundle over the real surface  
manifold. In other words one can consider such a surface either as a  
real manifold with a $U(1)$ bundle structure over it , or one may  
consider it as a complex manifold with a constant almost complex  
structure.

\bigskip
As a reasonable basis for the geometric quantization of our system,  
which defines a geometric quantization by a polarization of the  
underlying symplectic manifold of phase space \cite{wood}, our  
arguments are as follows: The reason that a "spatial" quantization  
on $M$ is possible is that in accord with Hirzebruch-Riemann-Roch  
theorem on a compact complex manifold without boundary $M^{\mathbf  
C}$, the Euler characteristic is given by $\chi := \int  
\limits_{\displaystyle{M^{\mathbf C}}} \bar{R} (TM) = \int  
\limits_{\displaystyle{M^{\mathbf C}}} ch_1 (TM^+)$ where $ch_1 (F)  
= c_1 (F) = (- 2 \pi i) ^{-1} F$ \cite{nak}. Here we consider only  
the case where $M$ is the above discussed two dimensional symplectic  
manifold and $F$ is a $U(1)$ valued Yang-Mills curvaure. Thus the  
surface integral of curvature two form on the complexified manifold  
$M$ is equal to the integral of first Chern class of a flat $U(1)$  
bundle over the complexified $M$, since $ch_1 (TM^+) = ch_1 (TM ;  
U(1)_{flat})$. Note that the geometric quantization of this system  
requires by the integrality condition that just the integral of  
$ch_1 (TM^+)$ on $M$ should be an integer multiple of Planck  
constant.

Recall also that the polarization, which is required by the  
geometric quantization \cite{wood}, is here due to the holomorphic  
polarization on $TM$ which is given by considering the Chern class  
on $ TM^+$.

Moreover the required complexness of the underlying manifold by the  
Hirzebruch-Riemann-Roch theorem is given here by the above  
mentioned property that the two dimensional manifold $M$ is a real  
manifold with vanishing Nijenhuis tensor. Hence it admits a global  
complex structure which simulates the global $U(1)$ bundle structure  
which is required to define the first Chern class: $c_1 (F)$.  
Therefore all condition which are needed for a geometric  
quantization can be fulfilled in the above discussed case of $(M ;  
\bar{R})$.

In other words, in view of the fact that Euler characteristic on a  
two dimensional orientable manifold without boundary $M$ is always  
given by $\chi := \int \limits_M \bar{R} (TM)$ and in accord with  
the above theorem that $\chi := \int \limits_M \bar{R} (TM) = \int  
\limits_M ch_1 (TM ; U(1)_{flat})$.
Then the quantization which is achived by the integrality of first  
Chern class of a flat $U(1)$ bundle over this two dimensional  
manifold, i. e. by $\int\limits_M  ch_1 (TM ; U(1)_{flat}) = N h$,  
can be expressed  by the integrality of the equivalent Euler class,  
i. e. by $\int \limits_M \bar{R} (TM) = \int\limits_{(surface)}  
R_{mn} dx^m \wedge dx^n = N h$. Also since the curvature tensor  
$R^{(surface)} _{mn} = \displaystyle{\frac{1}{2}}(\partial_m  
\Gamma_n - \partial_n \Gamma_m )$ on the surface $(M ; \bar{R} )$  
can be considered as the $U(1)$ curvature, i. e. as the field  
strength $F_{mn} = \displaystyle{\frac{1}{2}} (\partial_m A_n -  
\partial_n A_m )$ on the simulated  $U(1)$ bundle over $M$. This  
underlines the application of the above introduced theorem in our  
case with respect to the relation between the Euler class $( \sim  
\bar{R} = R_{mn} dx^m \wedge dx^n)$ and Chern class $( \sim F =  
F_{mn} dx^m \wedge dx^n)$.

Thus for relevant physical conditions where the two dimensional  
gravitational $W_G$ system has to be considered as a quantum system,  
the canonical quantization of this system can be described in  
accord with the geometric quantization by the quantum postulate $W_G  
= \int\limits_M \bar{R} = N h$.

\bigskip
A physical example of such a quantization on a two dimensional  
configuration space is the cyclotron motion of electron in a  
magnetic field \cite{aok} \cite{erkkme}. In this two dimensional  
case the canonical quantization is given similar to the  
experimentally varified flux quantization by postulating: $ \int  
\limits_{surface} e \tilde{F}_{mn} dx^m \wedge dx^n = \oint  
\limits_{contour} e \tilde{A}_m dx^m = N h$ where $\tilde{A}$ and  
$\tilde{F}$ are the $U(1)$ valued electromagnetic $1$ and $2 forms$.  
Thus if one considers the flux quantization as the canonical  
quantization of a two dimensional electromagnetic system of  
electrons \cite{erkkme}, then the equivalent quantum commutator  
postulate is given by $e B [ \hat{x}_m , \hat{x}_n ] = -i \hbar  
\epsilon_{mn}$ which is known, {\it phenomenologically}, as the  
non-commutativity of the relative coordinate operators $\hat{x}_m$  
in the cyclotron motion of electrons \cite{aok}, where $B$ is the  
applied spatially constant magnetic field. There is also an  
uncertainty relation $e B \Delta x_m \cdot \Delta x_n  
|\epsilon_{mn}| = e \Delta A_m \cdot \Delta x_m \geq \hbar$  
\cite{erkkme}.

Obviously this two dimensional electromagnetic quantization is  
comparable with the above quantization of two dimensional gravity.  
Thus the magneto-quantization on the two dimensional system is due  
to the applied spatially constant strong  magnetic field $F$,  
whereas the gravito-quantization is due to the spatially constant  
strong gravitational field $R$ of BH.

\bigskip
A geometrical analysis of this quantum structure shows that, as it  
is mentioned above, the curvature tensors $F_{mn}$ or $R_{mn}$ act  
here as {\it constant} almost complex structres on respective two  
dimensional manifolds which provide the underlying two dimensional  
manifolds with the flat $U(1)$ structure that is necessary for the  
geometric quantization of respective action functionals. Hence it is  
the values of these field strengths and of areas of respective  
intercations regions that determine the value of action functional  
which decides about the dominance of quantum or classical level of  
intercation. Recall that in accord with \cite{feyn} in the case  
where $W \sim \hbar$ the quantum modes are dominant.

However, whereas the two dimensional electromagnetic quantization  
is verifiable in strong magnetic fields by cyclotron motion or flux  
quantization, the two dimensional gravitational quantization is, in  
view of the $10^{-40}$ weakness of gravitational interaction with  
respect to electromagnetic interaction, only verifiable in very  
strong gravitational field of BH.

\bigskip
Footnotes and references


\begin{thebibliography}{100}


\bibitem{haw}
S. W. Hawking, Commun. Math. Phys., 43, 199-220 (1975);

See also the metric approach to the original model, e. g.: R. M.  
Wald, "General Relativity" (The University of Chicago Press 1984).



\bibitem{a}
The invariant or inherent dimensions of curvature- or field  
strength two-form can be determined in accord with the following  
fact that every r-form is dimensionless. Hence in view of the fact  
that $dx^i \sim x^i \sim L$, then $dx^i \wedge dx^j \sim L^2$ and  
the $R_{ij}$ tensor components of the two-form $R_{ij} dx^i \wedge  
dx^j$ are of dimension $L^{-2}$. In the same manner a potential  
component of a connection one-form is of invariant dimension  
$L^{-1}$. Note that this invariant dimension is different than the  
usual dimension which vary with the dimension of the underlying  
manifold where the forms are defined.


\bibitem{KAM}
V. I. Arnol'd: "Mathematical methodes in classical mechanics",  
(graduate text in mathematics, Springer-Verlag 1978). Note that only  
the cases with $n =2$, or those which can be reduced to them, have  
a satisfactory solution.

\bibitem{nnnn}
The precession of gyroscope is, like the Lense-Thiring effect,  
given by the {\it angular} velocity $\dot{\phi}$ which is in accord  
to the definition of $\phi$ also a two dimensional effect, since  
time should be considered in the phase space of system, not as a  
variable, but only as a parameter.

\bibitem{wit2}
E. Witten, hep-th/9206069.

\bibitem{nnn}
Recall also that even the electromagnetic waves ($\sim$ photon )  
possess only two degrees of freedom.

\bibitem{kr}
Note that the $\theta$ and $\phi$ "coordinates" in this metric are  
spherical coordinates in subspaces with spherical symmetry, so that  
they can not be considered as true independent coordinates and can  
be suppresed with respect to the analysis of singularities.


\bibitem{wood}
N. Woodhouse, "Geometric Quantization", (Oxford University,  
Clarendon Press, 1980, 1990); N. Hitchin, Commun. Math. Phys., 131,  
347-380 (1990).

Note with respect to the quantization that the question of  
metaplectic or corrected geometric quantization which founds the  
usual $(\displaystyle{\frac{1}{2}})$ term in the energy quantization  
postulate $( \sim W = \displaystyle{\frac{E}{\nu}}:= (N +  
\displaystyle{\frac{1}{2}}) h)$ is here omited (see the above cited  
Book).

\bibitem{typ}
Typical global or topological quantum effects are Aharonov-Bohm  
effect, Flux quantization and quantum Hall effect, which are caused  
by the flux of magnetic field through a surface. The electrons  
moving on the contour of such a surface perceive a phase change  
which is given in accord with the Stokes' theorem by: $e  
\int_{(surface)} F_{mn} dx^m \wedge dx^n = e \oint_{(contour)} A_m d  
x^m $ where $e$, $F$ and $A$ are the the electric charge, magnetic  
field and the electromagnetic potential. Obviously these results are  
two dimensional topological invariants in the sense that the flux  
surface is a two dimensional manifold.

\bibitem{aoki}
The magnetic length is defined by $L_B ^2 =  
\displaystyle{\frac{\hbar}{e B}}$ where $B$ is the applied constant  
magnetic field which is the curvature of the $U(1)$ bundle. For  
phenomenological definition of magnetic length see text books in  
solid state physics.


\bibitem{NN}
Recall that the quantization structure for the $(3+1)$ dimensional  
$U(1)$ Yang-Mills theory:

$\int\limits_{(3+1)} F \wedge * F$ is postulated by $[ \hat{A}_i \,  
, \hat{E}_j ] \propto \hbar \delta_{ij}$ where $E_i := \dot{A}_i$  
and $i, j = 1, 2, 3$, whereas the quantization of the $(2+1)$  
dimensional $U(1)$ Chern-Simons theory: $\int\limits_{(2+1)} A  
\wedge F$ is postulated in accord with
$A_0 =0$ gauge fixing by $[ \hat{A}_m \, , \hat{A}_n ] \propto  
\hbar \epsilon_{mn}$ where $m, n = 1, 2$ (see also Ref. [3]).

\bibitem{feyn}
R. P. Feynman, A. R. Hibbs, "Quantum mechanics and path integrals"  
(Mc Graw-Hill 1965).





\bibitem{nn}
Of course one can obtain the same equations of motion also from the  
equivalent action functional on the contour region $W_G =  
\oint\limits_{contour} \Gamma_m dx^m$. However in this case, i. e.  
on the contour region, one should consider that the equation of  
motion $\partial_m \Gamma_m = 0$ or $d^{\dagger} \Gamma = 0$ should  
be fulfilled by flat connection $\Gamma_{contour} := \Gamma_{flat}$  
for which $\epsilon_{mn} \partial_n \Gamma_m = 0$ or $d \Gamma = 0$.  
These two group of equations, i. e. $d^{\dagger} \Gamma = 0$ and
$d \Gamma = 0$ together are equivalent to the same Laplace equation  
$(d^{\dagger} d + d d^{\dagger}) \Gamma = \partial_n \partial^n  
\Gamma_m = 0$.



\bibitem{erkBAme}
This is a typical situation in topological quantum effects, as e.  
g. in Bohm-Aharonov effect, where just the contour integral of the  
locally vanishing flat connection, which is equal to the topological  
surface integral of non-vanishing curvature, causes the observable  
quantum phase.



\bibitem{quant}
According to geometric quantization \cite{wood} the classical  
vector fields related to the canonical conjugate variables ${\{  
\pi_m \, ,\, q^m }\}$ on the phase space are given by:

$X_{\pi_m} = \displaystyle{{\frac{\partial \pi_m}{\partial  
\pi_n}}{\frac{\partial}{\partial q^n}} - {\frac{\partial  
\pi_m}{\partial q^n}}{\frac{\partial}{\partial \pi_n}}}$ \qquad,  
\qquad
$X_{q^m} = \displaystyle{{\frac{\partial q^m}{\partial  
\pi_n}}{\frac{\partial}{\partial q^n}} - {\frac{\partial  
q^m}{\partial q^n}}{\frac{\partial}{\partial \pi_n}}}$

Furthermore, the inner product of any globally hamiltonian  
vectorfield $X_f$ of a function $f$ on the phase space of system  
with the symplectic 2-form $\omega = d\pi_m \wedge dq^m$, should  
result in: $< X_f , \omega > =  df$. In other words $X_{\pi_m} =  
\displaystyle{\frac{\partial}{\partial q^n}}$ and $X_{q^m} = -  
\displaystyle{\frac{\partial}{\partial \pi_n}}$.

Moreover in the polarized (quantum) phase space, i. e. in a certain  
representations as in $\Psi (\pi_m , t)$- or $\Psi (q^m , t)$  
representation, the quantum operators for ${\{ \pi_m \, ,\, q^m }\}$  
variables are given, respectively, by:
${\{ \hat{\pi}_m = \pi_m \, , \hat{q}^m = -i \hbar X_{q^m} = i  
\hbar \displaystyle{\frac{\partial}{\partial \pi_n}} }\}$ or ${\{  
\hat{\pi}_m = -i \hbar X_{\pi_m} = -i \hbar  
\displaystyle{\frac{\partial}{\partial q^n}} \, ,\, \hat{q}^m = q^m  
}\}$.


\bibitem{x}
For a four dimensional model of "quantum space time" and the  
discussion of other four dimensional models see: S. Doplicher, K.  
Fredenhagen, J. E. Roberts, Commun. Math. Phys., 188-220 (1995).

Note that the central question of space-time quantization, i. e.  
the physical meaning of the commutator postulate in this model, is  
not answered.

\bibitem{nak}
M. Nakahara: "Geometry, Topology And Physics" (Adam Hilger, 1990);
C. Nash: "Differential Topology and Quantum Field Theory",  
(Academic Press 1991).



\bibitem{aok}
H. Aoki, Rep. Prog. Phys, 50, 655 (1987).


\bibitem{erkkme}
For a general discussion of canonical quantization of two  
dimensional electrodynamics see:

F. Ghaboussi: (cond-mat/9710092), (quant-ph/9702054),  
(cond-math/9701128), (cond-math/9703080).



\end{thebibliography}
\end{document}